**All-electric picosecond field-free spin-orbit torque switching in magnetic trilayers**

Xinhou Chen, Shishun Zhao, Yuchen Pu, Qu Yang and Hyunsoo Yang*

Department of Electrical and Computer Engineering, National University of Singapore, Singapore 117583

*e-mail: eleyang@nus.edu.sg

**Spin-orbit torque (SOT) enables the electrical manipulation of the magnetization with high speed and low energy consumption for magnetic random-access memory (MRAM) applications. Previous studies of short-pulse SOT switching have mainly focused on the nanosecond regime, whereas reports employing picosecond pulses remain scarce and have largely relied on field-assisted switching using bulky, high-power laser systems, limiting prospects for chip-level integration. Here, we introduce an all-electrical on-chip nanoplasma pulse generator capable of producing pulses as short as 6.4 ps, enabling ultrafast picosecond field-free SOT switching in magnetic trilayers. We show that reducing the pulse width lowers the writing energy by 2–3 orders of magnitude, with ultrafast Joule heating assistance playing an essential role in the enhanced efficiency of the picosecond regime. Our demonstration of ultrafast, all-electrical, and field-free SOT switching establishes the nanoplasma pulse generator as an on-chip platform for ultrafast spintronic studies, with promise for high-speed, energy-efficient, and scalable SOT-MRAM technologies.**

Spin-orbit torque[1,2] (SOT) significantly enhances durability and reduces the energy consumption of magnetic random-access memory (MRAM)[3-7]. Conventional spin-torque memory operation is limited to GHz speeds, while SOT shows promise for even higher-speed applications[8-13]. Achieving ultrafast SOT-MRAM with operation frequency beyond GHz necessitates reducing the width of writing current pulses below nanoseconds even down to picoseconds. Recently, laser-driven Auston switches have enabled ultrafast demagnetization in Ta/GdFeCo/Pt and field-assisted SOT switching in Pt/Co/Ta using pulses shorter than 10 ps[14-17]. However, femtosecond laser amplifiers are required to generate picosecond current pulses, which presents substantial challenges for integration on chips. This underscores the necessity for developing ultrafast, all-electric, and field-free SOT switching with picosecond pulses.

To achieve ultrafast field-free SOT switching, a pulse generator capable of producing picosecond pulses with high amplitudes is required. However, it is challenging for conventional electric devices to achieve a rising rate of 1 V/ps due to the low Johnson figure-of-merit (JFOM)[18,19]. To address the narrow pulse generation issue, we introduce an all-electric pulse generator based on a nanogap, capable of producing pulses as short as 6.4 ps with amplitudes exceeding 100 V. In addition, we integrate magnetic trilayers consisting of Co/Pt/Co[20], which can realize field-free SOT switching due to the out-of-plane spin polarization (*z*-spins). Among various field-free SOT switching schemes,[21-32] the trilayer scheme emerges as a promising approach due to its excellent compatibility with existing industry fabrication processes. We demonstrate field-free SOT switching with pulse widths as short as 6.4 ps, and the writing energy density is four orders of magnitude lower than that achieved with 100 μs pulses. This work establishes an all-electrical route to ultrafast field-free SOT switching and provides a platform for high-speed spintronic devices and future SOT-based memory and logic applications.

**Nanoplasma based picosecond pulse generator**

Employing nanoplasma[19], we integrate a ps-pulse generator with a SOT device that consists of a coplanar waveguide, resistor, delay line, and nanogap, as depicted in Fig. 1a. The GaN switching circuit, as shown in Fig. 1b, generates a driving pulse with a few nanoseconds rise time. The driven pulse is then injected into the coplanar waveguide via the RF probe at contact A to trigger nanoplasma at the nanogap (Fig. 1c for illustration, Fig. 1a inset for scanning electron microscope image). Before the voltage of the driven pulse reaches the threshold $V_{\mathrm{TH}}$, the nanoplasma switch remains in the off state, with an infinite resistance of the nanogap, and thus the entire applied voltage drops across the nanogap. During this process, the delay line accumulates and stores energy, as depicted in Fig 1d. Once the applied voltage reaches $V_{\mathrm{TH}}$, nanoplasma initiates at the nanogap and rapidly transforms the nanogap from an insulating to a conducting state. As conducting plasma in the nanogap exhibits a very low resistance, the driving voltage is predominantly dissipated across the resistor rather than the nanogap. Consequently, the driving signal is isolated by the resistor and no longer provides energy to sustain plasma. Only the energy stored in the delay line sustains plasma and is injected into the SOT device. Once the energy stored in the delay line dissipates and the nanogap cannot sustain the plasma, the nanogap abruptly reverts to the insulating state, finishing the pulse generation process. To generate a negative pulse, a symmetric structure is fabricated on the opposite side of the SOT device. In this case, the driving voltage pulse is applied at contact D, and contact B is terminated.

To probe the pulse waveform, a 113 GHz oscilloscope is connected to contact *B*, while the driving source is connected to contact A, as illustrated in Fig. 1a and 1b. The pulse generator with an 800 nm gap and a 200 μm delay line produces a 6.4 ps pulse with an amplitude of 100 V as shown in Fig. 1e. Notably, its switching speed of 21 V/ps surpasses that of commercial GaAs/InP (1 V/ps)[19,33]. As illustrated in Fig. 1f-k, both the pulse width and amplitude can be tuned by modifying the length of the delay line and the width of the nanogap, respectively. When the length of the delay line ranges from 10 mm to 100 μm, the pulse width varies from 0.2 ns to 6.4 ps. Further decreasing the length of the delay line could achieve even shorter pulses, however, a pulse width of less than 6.4 ps cannot be detected correctly due to the bandwidth limitation of the 113 GHz oscilloscope. A wider nanogap leads to a higher $V_{TH}$[19,34], which in turn charges the delay line more and produces pulses with larger amplitudes.

This all-electrical, compact, and integrable picosecond pulse generator can be further miniaturized and fabricated directly on silicon substrates (Supplementary Fig. 1a-c). Together with its tunable high-amplitude picosecond output, this establishes the nanoplasma generator as an on-chip platform for ultrafast spintronic studies and an alternative to laser-based picosecond excitation. With further layout optimization and heterogeneous integration, it may also provide a forward-looking route towards more compact ultrafast SOT-MRAM architectures (Supplementary Note 1).

**Field-free SOT switching of Co/Pt/Co trilayers**

For the SOT device, we deposit a Co (8 nm)/Ta (0.5 nm)/Pt (1.5 nm)/Co (0.92 nm)/MgO (2 nm)/$SiO_2$ (3 nm) stack on sapphire substrates. As illustrated in Fig. 2a, the spin source layer (Co/Ta/Pt) is patterned into a Hall bar, while the top magnetic layer (Co/MgO/$SiO_2$) is patterned into a pillar. The magnetization of the bottom Co layer ($M_{bot}$) is oriented along the in-plane direction, while that of the top Co layer ($M_{top}$) is along the out-of-plane direction, as illustrated in the inset of Fig. 2a. As depicted in Fig. 2b, the square-shaped anomalous Hall loop confirms perpendicular magnetic anisotropy (PMA) of the top Co layer. The charge current flowing in the bottom Co/Ta/Pt layer generates spin currents with the spin polarization components along the *y*-direction and the $M_{bot} \times \hat{y}$ direction[35-37]. Since $M_{bot}$ is parallel or anti-parallel to the *x*-direction, the $M_{bot} \times \hat{y}$ component effectively contributes to the *z*-spins, facilitating the field-free SOT switching of the top Co layer[38-40].

We first conduct switching measurements using 100 μs current pulses. The magnetization of the bottom Co layer ($M_{bot}$) is preset along the *x* direction using an external magnetic field of 90 mT and then the external field is removed. The field-free SOT switching behavior is observed with the critical switching current density ($J_c$) of $3.5\times10^7$ A/cm$^2$, as shown in Fig. 2c. Furthermore, when we reverse the initial direction of $M_{bot}$, the switching polarity is reversed, confirming that the field-free switching is attributed to the *z*-spins associated with the magnetization direction of the bottom Co layer. The switching current density decreases if an assist field is applied along the *x*-axis. Without the bottom Co layer, the SOT switching in a Pt/Co bilayer requires an assist field along the *x*-direction (Supplementary Note 2 and 3), suggesting an important role of the bottom Co layer for field-free switching.

We perform loop shift measurements by probing the anomalous Hall effect (AHE) at different current densities to further confirm the presence of *z*-spins (Supplementary

Note 4). As shown in Fig. 2d, when the current ($I_{DC}$) exceeds 7.5 mA, a noticeable AHE loop shift occurs with $M_{bot}$ preset to either $+x$ or $-x$ direction. The existence of the threshold current for the noticeable AHE loop shift confirms the presence of damping-like torque induced by $z$-spins[20,38], with an effective out-of-plane field of ~2.2 mT at $3\times10^7$ A/cm$^2$.

**Picosecond current pulse-induced field-free SOT switching**

Utilizing the nanoplasma pulse generator, we investigate ultrafast field-free SOT switching in the picosecond regime. Figure 3 shows the switching behaviors of the Co/Pt/Co device observed by polar magneto-optical Kerr effect (MOKE) microscopy after applying a single 6.4 ps pulse. When $M_{bot}$ is preset along the $+x$ direction, $M_{top}$ switches from $-z$ to $+z$ with the current along the $-x$ direction (Fig. 3a and 3b), and $M_{top}$ switches from $+z$ to $-z$ with the current along the $+x$ direction (Fig. 3e and 3f). The switching polarity is reversed when $M_{bot}$ is preset along the $-x$ direction as shown in Fig. 3c and 3d as well as Fig. 3g and 3e. The above switching polarities are consistent with that observed with 100 μs current pulses in Fig. 2c. By evaluating the average pixel grayscale in the switching area of the MOKE images, we estimate the switching ratio to be greater than 90%, and the switching ratio decreases with lowering amplitudes of injected currents (Supplementary Note 5). The Pt/Co control device also shows similar behaviors but with an external magnetic field (Supplementary Note 6). Once the device is switched, it remains stable even though subsequent pulses are applied (Supplementary Note 7 and 8). The deterministic switching behavior depends on the current direction and the magnetization direction of the bottom Co layer, indicating that the switching behavior is dominated by the SOT effect.

**Switching current density and energy density relationship with pulse width**

To study the $J_c$ across a wide range of pulse width ($\tau$), we utilize the commercial nanosecond pulse source in addition to our nanoplasma pulse generator for $\tau$ from 6.4 ps to 100 μs. The SOT device and related RF circuits are designed to maintain good impedance matching, as shown in Supplementary Note 9. Figure 4a presents the $J_c$ versus $\tau$ relationship for field-free SOT switching in the Co/Pt/Co trilayer. For comparison, conventional field-assisted SOT switching behavior from the Pt/Co bilayer is also studied (Supplementary Notes 11 and 12). The switching dynamics of Co/Pt/Co show two distinct regimes. In the long-duration regime ($\tau$ > 10 ns), $J_c$ follows a $-\log^{0.5}(\tau)$ dependence and exhibits only moderate variation with $\tau$, consistent with previous reports[8,9]. As $\tau$ decreases from 10 ns to 0.35 ns, the measured $J_c$ exhibits a $\tau^{-1}$ dependence. In this regime, switching requires the injection of sufficient angular momentum to rotate the magnetization onto the equatorial plane (i.e., the top of the energy barrier). Given that a fixed minimum amount of angular momentum is required, a decrease in $\tau$ requires an increase in $J_c$, resulting in a $J_c \propto \tau^{-1}$ dependence[9,41,42]. From a domain wall motion aspect, nanosecond pulse-induced switching begins with the nucleation of magnetic domains at the edges, which then expand via edge-to-edge domain wall motion[10,11,43]. Since the wall velocity is proportional to the applied current intensity[44], and a full reversal requires the domain wall to traverse the entire sample, a shorter $\tau$ demands a higher $J_c$, resulting in a $\tau^{-1}$ dependency[45].

The relationship between switching energy density and pulse width is shown in Fig. 4b. Here, the switching energy density $\zeta$ is defined as

$$\zeta = \frac{W}{V} \propto J_c^{\,2} \cdot \tau \tag{1}$$

where $W$ represents the energy cost per switching event, and $V$ denotes the volume of the top Co layer. In our results, $J_c$ exhibits a $\tau^{-1}$ dependence within the 0.35~10 ns pulse range, leading to $\zeta \propto \tau_0/\tau + \tau/\tau_0$, where $\tau_0$ is the characteristic timescale of switching dynamics.[41,42,46] According to this prediction, as $\tau$ decreases, $\zeta$ initially decreases, reaching a minimum at $\tau_0$, and then scales inversely for shorter pulse widths. However, as the pulse width decreases below 0.35 ns into the picosecond regime, we observe that $J_c$ deviates from the conventional $\tau^{-1}$ scaling, and better fits to $J_c \propto \tau^{-0.35}$. The deviation leads to the switching energy density of $\zeta \propto \tau^{0.3}$ for $\tau < 0.35$ ns, and causes $\zeta$ to decrease as the pulse width shortens. This reduction in energy consumption highlights the enhanced efficiency of SOT switching under picosecond pulses. The above energy scaling behavior is also observed in the Pt/Co system with an assisted field (Supplementary Note 12).

**Efficient switching mechanisms with ps-pulse**

The high switching efficiency observed with narrow pulses primarily arises from coupled ultrafast Joule heating. Due to the limited heat capacity of the ultrathin film, picosecond pulses can induce a rapid and intense local temperature rise. Finite element simulations (Supplementary Note 13) confirm that, at their respective critical currents, picosecond pulses raise a significantly higher peak sample temperature. This transient thermal spike suppresses the saturation magnetization ($M_s$) and perpendicular magnetic anisotropy ($K_z$) of the PMA Co layer, effectively lowering the switching energy barrier and reducing the energy cost of picosecond pulse-driven SOT switching (Supplementary Note 13 and 14). Consequently, the $J_c - \tau$ scaling relationship deviates from the standard $\tau^{-1}$ behavior to $\tau^{-0.35}$.

Furthermore, using a material with a lower Curie temperature ($T_C$) that is more sensitive to thermal fluctuations, such as a $Co_{80}B_{20}$/Gd bilayer, the scaling exponent is further reduced to $\tau^{-0.28}$ as shown in Fig. 4a,b. The high $T_C$ Co film follows the ideal $\tau^{-1}$ behavior across the 0.35-10 ns range, as the induced temperature rise is insufficient to significantly alter its magnetic properties. In contrast, for materials with a relatively low $T_C$, we anticipate that this reduced exponent will manifest even for $\tau \geq 0.3$ ns. To verify, we have utilized a CoGdB alloy in a sapphire//Ta (1 nm)/Pt (6 nm)/CoGdB (3 nm)/Ta (2 nm)/$SiO_2$ (3 nm) stack, where the Gd concentration tunes $T_C$ from > 470 K down to 300 K (Fig. 5a). Due to a low $T_C$, even the limited Joule heating generated by pulses exceeding 0.3 ns suppresses $M_s$ and $K_z$. Consequently, the SOT switching $J_c - \tau$ scaling in CoGdB deviates significantly from $\tau^{-1}$ to $\tau^{-0.22}$ with lower $T_C$ compositions yielding smaller exponents (Fig. 5b). This correlation provides further evidence that the deviation from standard scaling is driven by thermal assistance, validating Joule heating as an essential contributor to the energy-efficient switching.

The second reason stems from the dynamics of magnetization switching induced by picosecond pulses. Conventional pulses in the nanosecond range or longer induce multiple precession cycles during magnetization switching, as shown in the switching trajectory simulations (Supplementary Fig. 18). These additional precession cycles lead to damping-induced dissipation, causing significant spin angular momentum loss and energy waste. To compensate, additional electrons must be injected to sustain the switching process against damping, thereby increasing overall energy consumption. In contrast, when an intense picosecond pulse is applied, the SOT-induce torque quickly pushes the magnetization to pass through the equator, minimizing damping-induced dissipation and angular momentum loss during reversal, thereby enhancing the switching efficiency and reducing the switching energy density.

We perform macrospin and micromagnetic simulations to assess the role of Joule heating in the switching process. As shown in Fig. 4c,d, when Joule heating is excluded, the $J_c$ - $\tau$ relationship from both the macrospin simulations (triangles) and the micromagnetic simulations (pentagons) exhibit the conventional $1/\tau$ scaling expected from angular momentum conservation and do not reproduce the experimentally observed enhancement in energy efficiency. After thermal effects are incorporated, the simulated results (red circles, Fig. 4c,d) align closely with the experimental data (blue squares). This comparison indicates that ultrafast thermal assistance is an essential component of the picosecond switching process in our system.

As bit density increases, closely packed small sized magnetic bits become more susceptible to thermal fluctuation, requiring the use of materials with high magnetic anisotropy to maintain stability[47,48]. However, higher anisotropy also increases the switching energy barrier, making external assistance necessary for magnetization reversal[49]. Laser-based heat assisted magnetic reversal overcomes this limitation in hard disk drives[50,51] but is incompatible with random-access, on-chip memory architecture. For high-density SOT-MRAM, heat assistance needs to be generated electrically and locally. Therefore, picosecond electrical pulses that induce transient Joule heating provide an effective means to establish ultrafast thermal assisted SOT switching.

## Conclusions

We have developed an all-electrical, integrable nanoplasma pulse generator capable of generating high-amplitude picosecond pulses. The nanoplasma pulse generator is distinguished by its straightforward design, cost-effective fabrication, and its unique ability to generate pulses as short as ~6 ps with amplitudes exceeding 100 V. The pulse widths and amplitudes can be controlled across a wide range, outperforming conventional high-speed electric devices. This establishes the nanoplasma pulse generator as an on-chip platform for ultrafast spintronic studies. Using this platform, we demonstrate field-free SOT switching in magnetic trilayers with pulse widths down to 6.4 ps and show that the writing energy density continues to decrease as the pulse width is reduced. We further show that ultrafast Joule heating assistance is an important reason for making the switching energetically favorable at the picosecond timescale. This work establishes an all-electrical route to ultrafast field-free SOT switching and provides a platform for high-speed spintronic devices and future SOT-based memory and logic applications.

# Methods

**Fabrication of nanoplasma pulse generator.** The nanoplasma pulse generator was fabricated on (0001)-sapphire substrates. Initially, a 4-inch sapphire wafer was cleaned, followed by the deposition of a 300 nm thick gold layer on top of a 13 nm thick titanium layer using an electron beam evaporator. The wafer was then segmented into 15 mm×15 mm smaller pieces. The nanogap, coplanar waveguide, and resistor structures were fabricated using EBPG 5200 electron-beam lithography (EBL) followed by ion beam etching. In detail, ZEP 520A resist was utilized at a coating speed of 3000 rpm and baked at 180 °C for 5 min. The EBL step employed a dose of 260 μC/cm$^2$, and then the exposed sample was developed in amyl acetate for 2 min and rinsed in a 9:1 methyl isobutyl ketone/isopropyl alcohol solution for 1 min. Subsequent photolithography used the PFI resist at a spin speed of 6000 rpm for 1 min, followed by ion beam etching to form the radiofrequency probe pads. Another photolithography step and a gold wet etch process were employed to create a window over the resistor line and to remove the gold layer on top of the titanium, forming a 10 kΩ resistor. A 12 nm thick layer of $SiO_2$ was then sputtered on top of the titanium line to prevent oxidation of the resistor.

The magnetic PMA film was then grown by magnetron sputtering in an ultra-high vacuum chamber with a base pressure below $2\times10^{-9}$ Torr, for a stack of sapphire/Ta (2 nm)/Co (8 nm)/Ta (0.5 nm)/Pt (1.5 nm)/Co (0.92 nm)/MgO (2 nm)/$SiO_2$ (3 nm)/Ta (0.25 nm). Ta served as a buffer layer, and a thick bottom Co layer was employed to decrease device resistance. Finally, the film was patterned into a microstrip through lithography and etching processes.

**Measurement of ultra-fast picosecond pulse.** The excitation signal for the nanoplasma switch was generated through a circuit configuration comprised of a 10 mH inductor in series with a 600 V/70 mΩ LMG 3410R070 GaN power transistor. A square waveform generator was employed to toggle the GaN transistor between open and closed states. When the square waveform is at a high level, the GaN transistor is open, allowing currents to flow through the inductor. Conversely, when the control waveform changes to a low level, the GaN transistor closes abruptly. This rapid transition forces the inductor to maintain the current flow, thereby generating a nanosecond pulse. This pulse serves as the driving signal for the nanoplasma switch. The generated pulse signal was captured using a 110 GHz probe and routed through a series of 110 GHz coaxial attenuators, cumulatively providing 40 dB of attenuation. The attenuated signal was then measured by a Keysight UXR1102A 113 GHz real-time oscilloscope. The power transfer efficiency between the threshold driver signal and the generated pulse amplitude is shown in Supplementary Note 15.

**De-embedding process.** To eliminate the impact of the measurement components before the oscilloscope, such as attenuators, a radiofrequency probe, a 150 mm long coaxial cable with 1.0 mm connectors, and two 1.0 mm converters, we employed the *S*-parameters provided by the component manufacturers. To synthesize the overall *S*-parameter matrix for the measurement chain, we first converted the *S*-parameter matrices into *ABCD* matrices. The formatted *ABCD* matrices were then multiplied sequentially to form a cumulative *ABCD* matrix for the entire chain. This comprehensive *ABCD* matrix was subsequently converted back into an *S*-parameter matrix, where $S_{21}$ serves as the transfer function for the entire measurement setup. The generated signal is extracted utilizing the inverse Fourier transformation.

**Picosecond pulse switching and MOKE imaging.** The image was captured using a MOKE microscope from Tuotuo Technology equipped with a 633-nm light source and an 80× objective lens. The RF circuit setup comprised a pair of 1.0 mm connector probe tips equipped with micro-positioners. One probe is connected to the driven source, while the other is linked to a 50 Ω terminal.

**Sub-nanosecond to microsecond pulse switching.** A PSPL 10060A pulse generator was used to create pulses ranging from 0.1 ns to 10 ns. For longer pulse durations, a Keysight 81134A pulse pattern generator was utilized. A 1 µm width channel is used to obtain enough current density due to the voltage limitation of the ns pulse generator. Before pulse application, the magnetization of the bottom Co layer in the stack was saturated in either the +*x* or -*x* direction, and the top Co PMA layer was initialized to an up or down magnetic state. Following each pulse injection, the anomalous Hall effect resistance ($R_{AHE}$) was measured using a 100 µs current to detect the state of the top Co layer, by operating

Keithley 6221 and 2182A in the pulse delta mode with 30 times average. After detecting that the SOT device was switched, it was reinitialized to its original state, and a different amplitude pulse was applied. The above iterative approach allowed for precise characterization of the switching dynamics under various pulse conditions.

After obtaining the critical pulse width and amplitude values, the actual pulse width $\tau$ and amplitudes $V_{\mathrm{osc}}$ after passing through the bias-tee and cables were calibrated using an oscilloscope. The real voltage applied to the device was evaluated using the formula $V_p = V_{\mathrm{osc}} \cdot 2R/(R + 50\,\Omega)$,[54] where $R$ is the resistor of the device. The equivalent critical current density was then calculated based on the geometry and resistance of the device.

**Data availability**

The data that support the plots within this paper and other findings of this study are available from the corresponding author upon reasonable request.

**Code availability**

The code used in this paper is available from the corresponding author upon reasonable request.

**Acknowledgments:** We are grateful to Keysight for providing the UXR1102A oscilloscope. The work is supported by National Research Foundation (NRF) Singapore Investigatorship (NRFI06-2020-0015), the Ministry of Education, Singapore, under Tier 2 (T2EP50124-0017), and Samsung Electronics Co., Ltd (IO241218-11518-01).

**Author contribution:** X.C. performed the picosecond pulse generator design. X.C. and S.Z. performed film depositions, device fabrications, and nanosecond measurements. Q.Y. helped with nanosecond measurements. X.C. and Y.P. performed DC measurements. X.C., S.Z. and Y.P. carried out simulations. X.C., S.Z., Y.P. and H.Y. wrote the manuscript. The project was initiated by X.C. and H.Y. All authors contributed to the discussions of the results.

## Competing interests

The authors declare no competing interests.

## Supplementary Information

The online version contains supplementary material available.

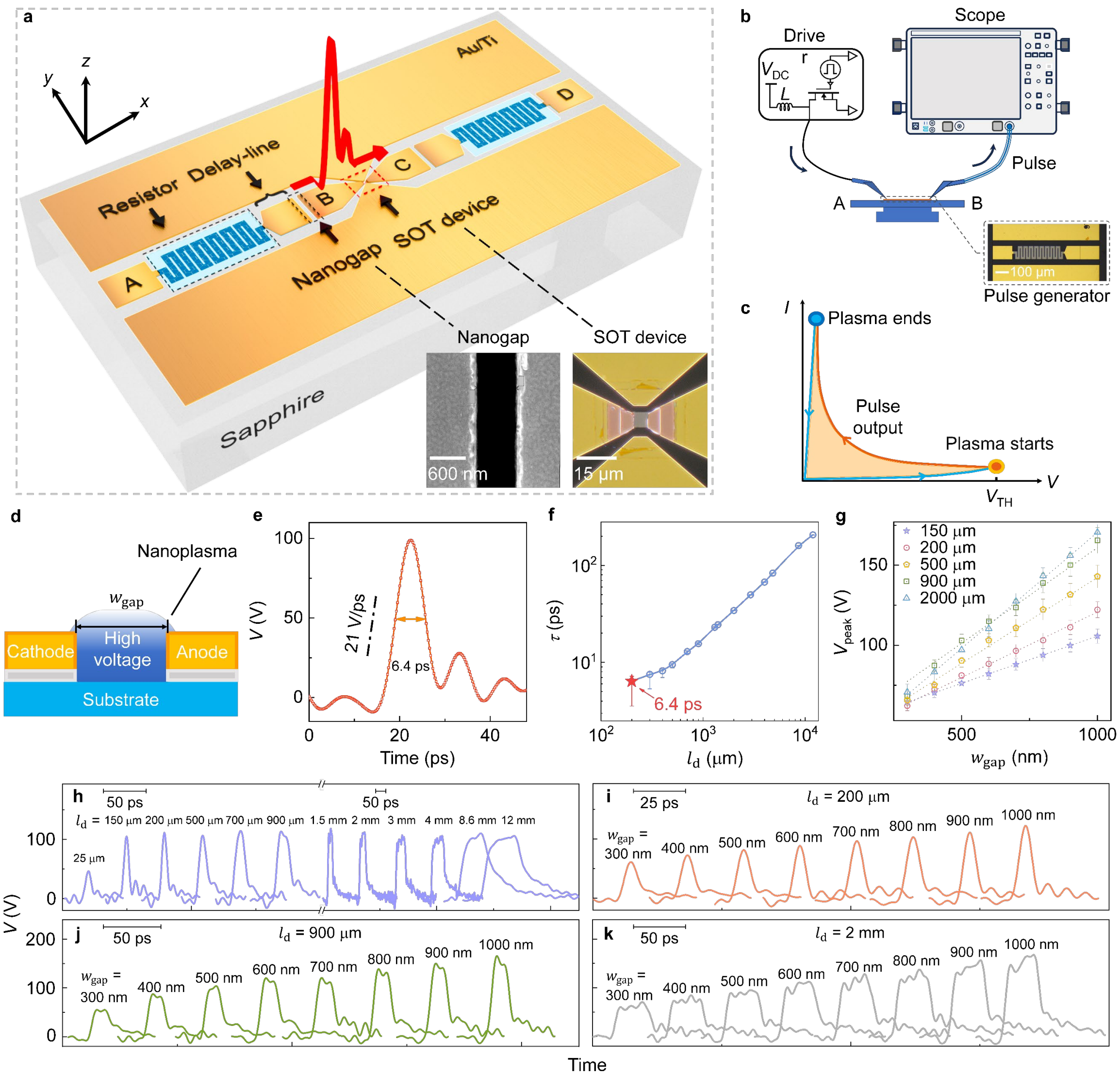

**Fig. 1 | Nanoplasma picosecond pulse generator for ultrafast SOT switching. a**, Overview of the nanoplasma pulse generator and the SOT device. Pads A and D are connected to driving sources, while pads B and C are terminated. **b**, Schematic of the platform for pulse calibration. The inset shows the pulse generator used for waveform calibration **c**, State diagram of the nanoplasma pulse generator. Arrows indicate the pulse generation sequence. **d**, Detailed schematic of the nanoplasma switch. Once the driving signal reaches the threshold voltage of the nanoplasma switch ($V_{TH}$), plasma is triggered, and field-emission electrons rapidly transport across the nanogap due to the high electric field. **e**, The pulse generator produces a minimum 6.4 ps high-amplitude pulse with a slew rate of 21 V/ps. **f**, Relationship between the pulse width ($\tau$) and the delay line length ($l_d$). **g**, Peak pulse output voltage ($V_{peak}$) versus nanogap width ($w_{gap}$) for $l_d$ of 150 μm, 200 μm, 900 μm, and 2 mm. **h**, Output pulses for different delay-line lengths. **i**–**k**, Waveforms at fixed $l_d$ with $w_{gap}$ swept from 300 to 1000 nm in 100 nm steps, for $l_d$ =200 μm (**i**), 900 μm (**j**) and 2 mm (**k**).

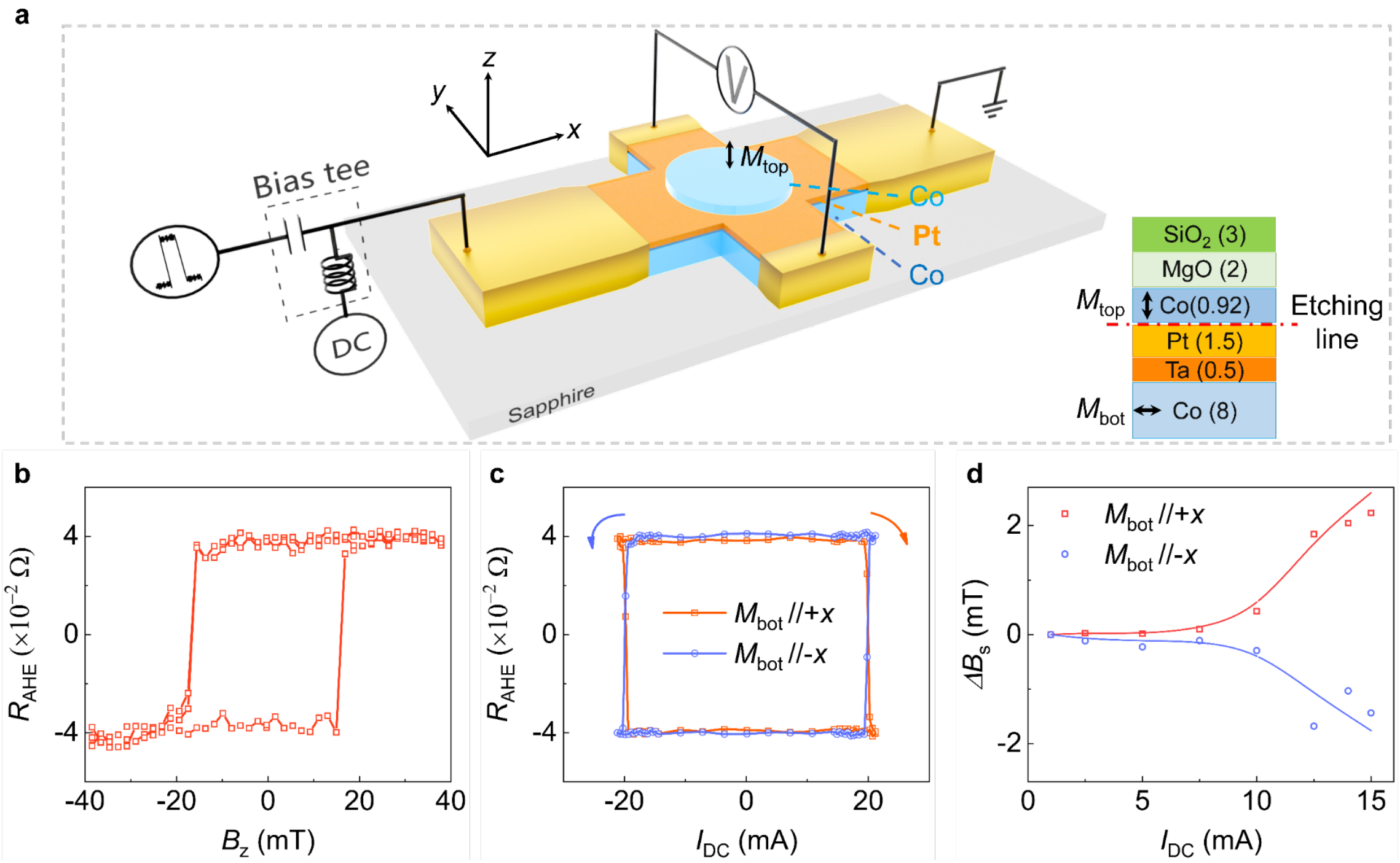

**Fig. 2 | Field free switching of Co/Pt/Co with 100 μs pulses. a**, Schematic for sub-nanosecond to microsecond pulse-induced switching measurements. $M_{top}$ and $M_{bot}$ refer to the magnetization of the top PMA Co layer and the bottom in-plane Co layer, respectively. **b**, The anomalous Hall effect of Co/Pt/Co devices with the out-of-plane magnetic field ($B_z$). **c**, Field-free SOT switching of Co/Pt/Co devices with the magnetization of the bottom Co layer ($M_{bot}$) preset along the +*x* or -*x* direction. **d**, The loop shift ($\Delta B_s$) obtained from the anomalous Hall measurements as a function of applied current ($I_{DC}$) with $M_{bot}$//+*x* or $M_{bot}$//–*x*. The $I_{DC}$ of 10 mA corresponds to the current density ($J_c$) of $1.76\times10^7$ A/cm$^2$ in the bottom Co/Pt layer.

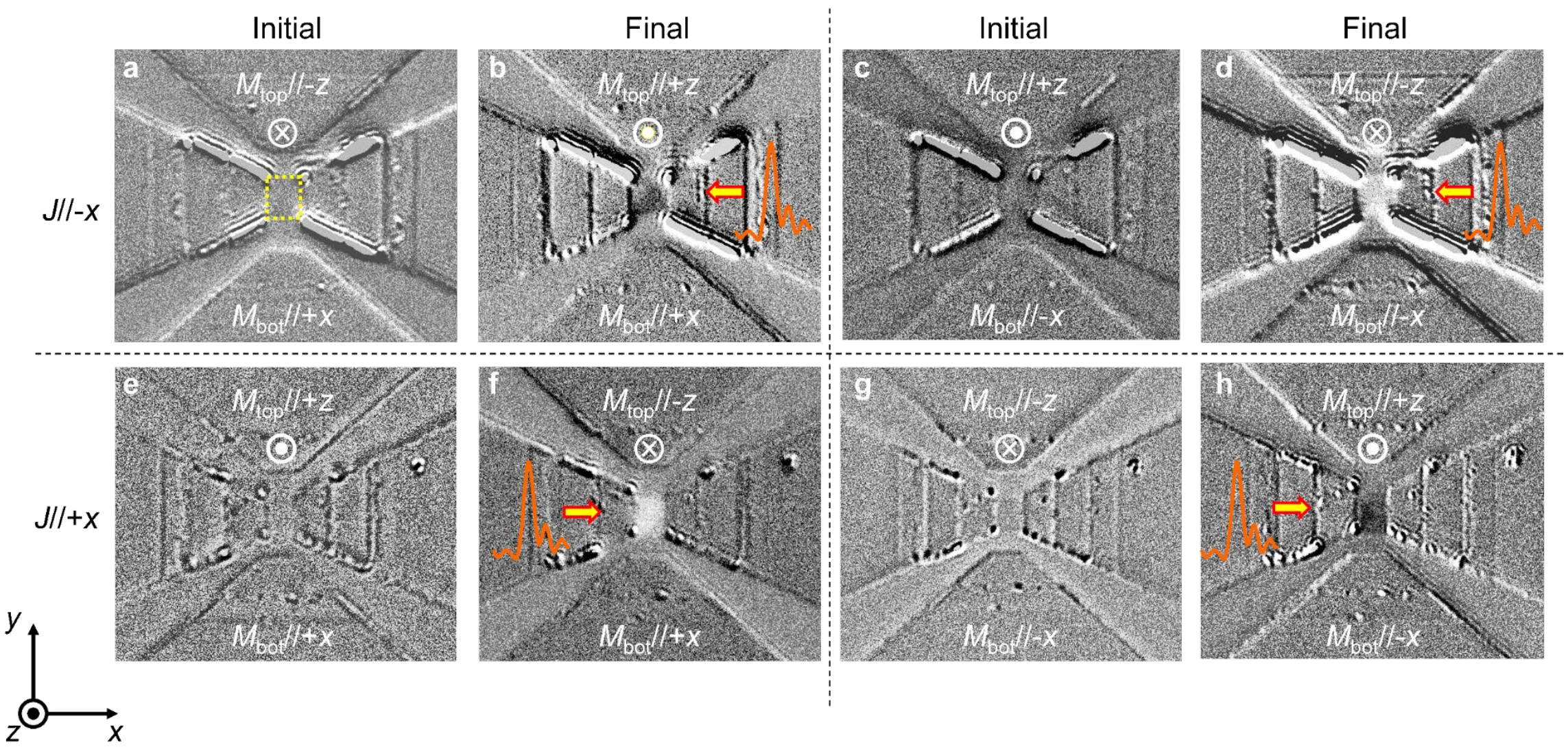

**Fig. 3 | MOKE images of Co/Pt/Co devices by applying a 6.4 ps pulse.** The ps-pulse is produced by the nanoplasma pulse generator with a 200 μm delay line and an 800 nm nanogap. The yellow dashed square in **a** indicates the Co/Pt/Co area. The four quadrants depict the initial and after-pulse domain state under different initial $M_{bot}$ and pulse directions. In **a**,**b**,**e**, and **f**, $M_{bot}//+x$; in **c**,**d**,**g**, and **h**, $M_{bot}//–x$. The dark and light areas represent the magnetization along the $+z$ and $–z$ direction, respectively.

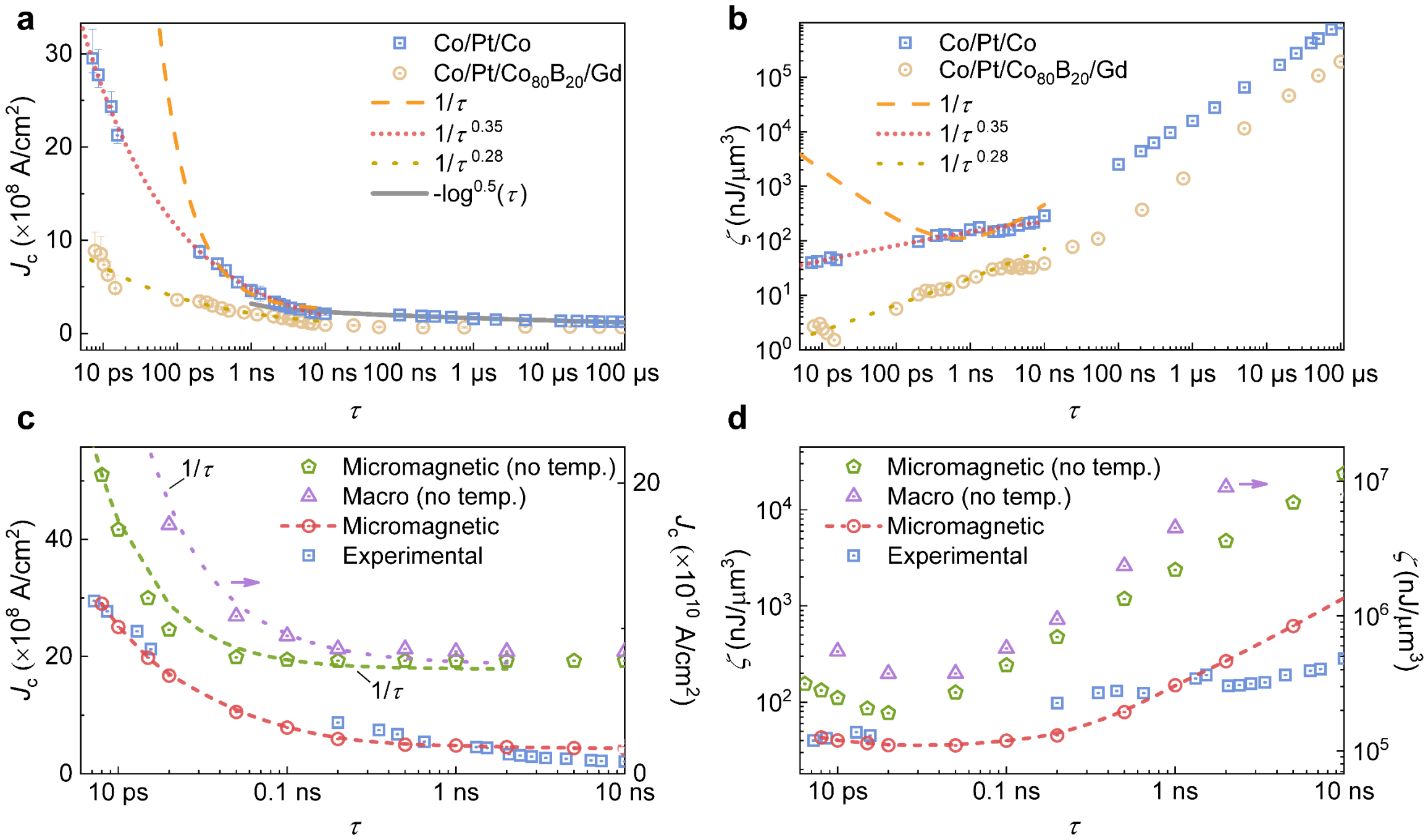

**Fig. 4 | Critical current density and energy dependence on pulse width for field-free switching in ferro- and ferrimagnetic films. a**, Critical switching current density ($J_c$) as a function of pulse width ($\tau$). For the longer pulse regime (> 10 ns), the grey line indicates the -$\log^{0.5}(\tau)$ fitting results. In the short pulse regime (< 10 ns), the orange dashed line represents a fit of $\tau^{-1}$, while the red dotted line is obtained with $\tau^{-0.35}$ function. Error bars reflect uncertainties arising from pulse calibration. **b**, Switching energy density ($\zeta$) as a function of $\tau$. The dashed and dotted lines correspond to the respective fits. **c**, Enlarged $J_c$ versus $\tau$ for $\tau < 10$ ns. **d**, $\zeta$ as a function of $\tau$ for $\tau < 10$ ns. Both panels compare results with no thermal effects (labeled no temp.) and thermal effects.

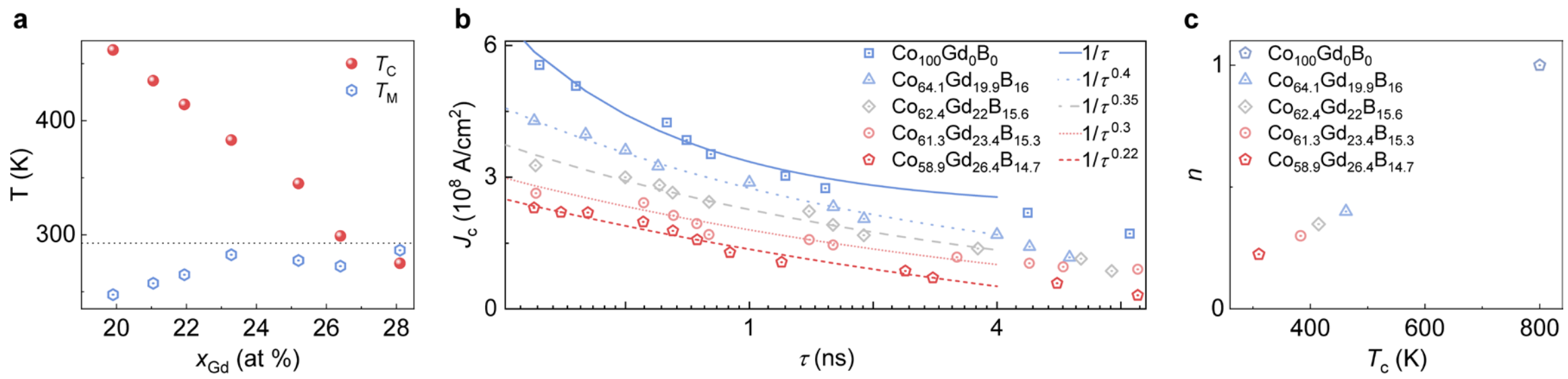

**Fig. 5 | Thermally assisted switching under short pulses. a**, Curie temperature ($T_C$, red) and magnetization compensation temperature ($T_M$, blue) of Co-Gd-B as a function of Gd concentration ($x_{Gd}$). The dashed line marks the sample temperature used for the subsequent SOT switching measurements. **b**, Critical switching current density $J_c$ versus pulse width $\tau$ for different $x_{Gd}$. Lines are fits to $J_c \propto 1/\tau^n$, where $n$ quantifies how rapidly $J_c$ increases as $\tau$ is reduced. **c**, Extracted exponent $n$ as a function of $T_C$. The relatively low $T_C$ of Co-Gd-B leads to reduced $n$ (slow increase of $J_c$ with decreasing $\tau$).